# Prioritizing Policies for Furthering Responsible Artificial Intelligence in the United States

Emily Hadley
*Center for Data Science*
*RTI International*
Durham, NC, USA
ORCID 0000-0003-0074-4344

*Abstract*— Several policy options exist, or have been proposed, to further responsible artificial intelligence (AI) development and deployment. Institutions, including U.S. government agencies, states, professional societies, and private and public sector businesses, are well positioned to implement these policies. However, given limited resources, not all policies can or should be equally prioritized. We define and review nine suggested policies for furthering responsible AI, rank each policy on potential use and impact, and recommend prioritization relative to each institution type. We find that pre-deployment audits and assessments and post-deployment accountability are likely to have the highest impact but also the highest barriers to adoption. We recommend that U.S. government agencies and companies highly prioritize development of pre-deployment audits and assessments, while the U.S. national legislature should highly prioritize post-deployment accountability. We suggest that U.S. government agencies and professional societies should highly prioritize policies that support responsible AI research and that states should highly prioritize support of responsible AI education. We propose that companies can highly prioritize involving community stakeholders in development efforts and supporting diversity in AI development. We advise lower levels of prioritization across institutions for AI ethics statements and databases of AI technologies or incidents. We recognize that no one policy will lead to responsible AI and instead advocate for strategic policy implementation across institutions.

## I. Introduction

Artificial intelligence (AI) technology has shown considerable promise for benefiting society, but irresponsible development and misuse have also led to great harm [1], [2]. Failures in AI technology have recently included the discovery of racial bias in healthcare algorithms [3], false arrests based on faulty facial recognition [4], and increasing concern about car crashes involving driverless or driver assistance tech [5]. Accountability for addressing these AI challenges lies not just with individual AI developers, but also with governments, agencies, educators, professional societies, and organizations that develop and deploy AI [6]. These institutional actors play a crucial role in implementing policies and practices that foster responsible AI.

AI is generally understood to be a machine or computer system able to perform tasks normally requiring human intelligence, including but not limited to making predictions, recommendations, or decisions [7]–[9]. Responsible AI considers the larger framework within which AI is developed and used and advocates for a value-driven process that prioritizes consideration of fairness, accountability, anti-discrimination, privacy, security, participatory engagement, explainability, sustainability, and societal impact [6], [10], [11]. In her book on responsible AI, Dignum [6] says that responsible AI is about, "ensuring that results are beneficial for many instead of a source of revenue for a few." Responsible AI is often presented together with ethical AI; however, responsible AI considers not only ethical concepts but also legal, economical, and cultural ones [6], [12]. Although ethical AI has been critiqued for making the "non-obvious" assumption that poor AI ethics and bad design alone produce harmful outcomes, responsible AI retains a focus on normative action and equity [13].

At its roots, AI technology is developed by individual developers, often computer scientists, data scientists, or software engineers. Although all individuals involved in AI development can and should take responsibility for potential implications of their work [14], an individual commitment to learn about and implement responsible AI is insufficient for scalable change. Peters [15] points out that developers are not, and should not, be expected to do the work of philosophers, psychologists, and sociologists; policies and practices should instead be enacted to support collaboration with experts to anticipate and mitigate risks as a standard of practice. AI developers often find that normative aspirations conflict with commercial values of efficiency, speed, and profit [16], [17]. Individual developers operate within this culture, generally with limited influence, and it is unrealistic to expect individuals alone to change these dynamics [10], [14]. Furthermore, AI incidents can rarely be traced back to a single team member or action, and responsibility instead lies with the entire network of AI actors, including organizations that develop and deploy AI and regulatory institutions [16].

Building a culture of responsible AI development and use requires implementing policy and practices throughout this network. Effective mechanisms exist at various levels, including

RTI International provided support for this research.

national and state legislatures, government agencies, professional societies, and private and public sector organizations [18]. National, state, and local legislatures often play a role in "hard governance" including laws and mandated policies [18]. Examples include the European Union (EU) General Data Protection Regulation (GDPR), which includes numerous data protection guidelines [19]; the California Consumer Privacy Act, which focuses on increased transparency for data subjects [20]; and the New York City Local Law 144 requiring bias assessments of AI tools used in hiring [21]. Government agencies, while often tasked with enforcing "hard governance" laws created by legislative bodies, are also often frequently involved in "soft governance" such as creation of rules and voluntary standards [22]. Professional societies are also involved in "soft governance," such as the IEEE Ethically Aligned Design Standards for AI systems [23]. Private and public sector entities, and academic institutions, can implement internal policies and procedures that support adherence with soft and hard governance approaches. Yet overall, the efforts across these organizations and institutions to date have been for narrow AI use cases, especially in the United States, and lack an explicit focus on responsible AI [6].

No single institutional practice or policy will be sufficient to ensure responsible AI; rather, all institutional AI actors in the network must be simultaneously engaged in developing and implementing policies and practices [1]. In Section II we review proposed and existing responsible AI policies. In Section III, we assess potential for use and impact in the United States and suggest prioritization for implementation by each institutional stakeholder. The prioritization is particularly novel because although several frameworks to support responsible AI have been developed [1], [9], [22], few provide a comprehensive, multi-stakeholder, ranked perspective of policy options in a format useful to institutional stakeholders. National and state government, agencies, professional societies, and organizations can use this resource to better understand where to focus responsible AI efforts and to appreciate the larger context in which their efforts must operate.

## II. Review Of Responsible AI Policies

This review considers nine policy and practice areas: licensure or certification of AI developers, AI ethics statements, pre-deployment assessments and audits, post-deployment accountability, databases of AI technologies or incidents, involvement of community stakeholders, policies that support responsible AI education, policies that support responsible AI research, and policies that support diversity in AI development. Policies were selected for evaluation if they met the following criteria: (1) they require implementation by an institution (government, organization, etc.) rather than an individual; (2) they were proposed in the responsible AI literature; and (3) given the immediate need for responsible AI, they could be realistically implemented in the near future.

### A. Licensure or Certification of AI Developers

Licensure or certification is used to recognize the expertise or competence of an individual and can be required prior to starting work in specific professions [24]. Licensure in the United States is often regulated by legislation at the state level and in many cases, a license may be required in every state where an individual practices. Certification is managed by a recognized nongovernmental authority in an area, such as a board or professional organization, and a certification recognizes competency in that specific area [24]. Both licensure and certification create a clear standard of expectations for individuals with the threat of professional sanctions if standards are not maintained.

The number of U.S. jobs requiring an occupational license is now almost 1 in 4 [25]. Mittelstadt [16] notes, "it is a regulatory oddity that we license professions providing a public service, but not the profession responsible for developing technical systems to augment or replace human expertise and decision-making within them." Examples of these industries include real estate agent [26], radiologic technician [27], and city bus driver [28]. Licensing the data scientists and AI developers working on these technologies could better develop public trust and ensure high-quality delivery of these critical services through standardization of expectations and knowledge.

Yet a look at the history of software engineering licensing paints a bleak picture of licensure for developers [16]. Texas became the first state to license software engineers in 1998. Licensure was not popular among practitioners [29] and the Association for Computing Machinery (ACM) explicitly opposed licensing of software engineers, considering it "premature" and unlikely to solve problems of software quality and reliability [24]. In 2013, the National Council of Examiners for Engineering and Surveying introduced the Principles and Practice of Engineer Software Engineering exam. This exam was discontinued in April 2019 after only 81 candidates took it across five administrations [30]. Although individuals who use the title *engineer* are generally required to be licensed in each state they practice to ensure that they have the specific education required to complete an engineering task, limited interest in licensing software engineers beyond Texas, robust exemptions to licensure even in Texas, and limited enforcement likely inhibited the success of software engineering licensure [31]. These experiences do not suggest a promising outcome for licensure of AI developers.

Certification that would promote responsible AI is similarly lacking in promise. One major issue is that the practice of data science, along with many career tracks in information science and technology, continues to be in a place of defining itself [32]. ACM and IEEE are two organizations that are gaining traction among AI developers, but they are large and include many careers tracks beyond the scope of AI. Both have published codes of ethics for members, but these have been described as "comparatively short, theoretical, and lacking grounded advice and specific behavioral norms"[16]. Furthermore, as Mittelstadt [16] describes, AI developers do not necessarily commit to public service and they do not serve the healthcare equivalent of a patient whose interests are granted primacy. ACM, IEEE, or a similar professional organization proposing a certification for responsible AI or data science will have to operate within a culture where much AI development happens in sectors that prioritize alignment with fiduciary duty toward shareholders rather than users and affected parties [16].

Yet even with these challenges, there remain calls for better professional benchmarking of data science qualifications [32].

In the responsible data science and AI space, Mittelstadt [16] remarks that this could likely look like targeting licensure or certification initiatives for developers creating tools with elevated risk or built for the public sector, such as facial recognition systems for policing. CertNexus and Udemy have proposed ethical AI certification programs, but these are short in scope and appear to have limited uptake [33], [34].

*B. AI Ethics Statement*

In the last 5 years, one of the most popular and publicized efforts to support the ethical or responsible development of AI and ML has been the drafting of AI Ethics frameworks, principles, guidelines, and statements by high-profile companies, organizations, and countries [15], [35]–[39]. Most AI ethics codes of conduct have originated with private companies and governmental agencies in the United States and the EU [7]. These publications are often drafted in response to public conversations on social and ethical issues surrounding AI and propose principles, tenets, or values central to AI development [10], [16]. Greene [13] says these statements "set the tone for conversations around ethics and AI/ML." Jobin [7] reflects on how the popularity of AI principles among a diverse set of stakeholders reflects both the need for ethical guidance and the strong interest in shaping it. The Artificial Intelligence Risk Management Playbook [40] recommends publishing or adopting AI principles to clarify an organization's values.

Common responsible AI themes in these publications included transparency, non-maleficence, responsibility, and privacy for the development and deployment of responsible AI [7]. Researchers studying these publications note a lack of discussion about other responsible AI themes, such as societal context for AI usage, social responsibility, and sustainability [2], [7]. Hagendorff [2] attributes this to "male-dominated justice ethics" resulting from a gender imbalance in the authors of these works such that the standards were "calculating" and "logic-oriented" rather than "empathic" or "emotion-oriented." An exception is the findings of the European Commission, which includes a recommendation of voluntary commitments related to, for example, environmental sustainability, accessibility for persons with disabilities, and stakeholder participation [1].

Although most AI ethics codes of conduct are focused on the behavior of institutional actors, a few approaches have also been suggested for individuals. Eubanks [41] proposes principles of non-harm, similar to the Hippocratic Oath, for individual data scientists, systems engineers, hackers, and administrative officials. This Oath of Non-Harm for an Age of Big Data includes tenets such as, "I will not use my technical knowledge to compound the disadvantage created by historic patterns of racism, classism, able-ism, sexism, homophobia, xenophobia, transphobia, religious intolerance, and other forms of oppression" [41]. Legal and technical scholars have called for regulating data scientists as fiduciaries [13], [42]. There is debate around whether it is more effective to focus on individual or group accountability. Barredo Arrieta [11] says that it is wiser to focus on the self-responsibility of an employee. Mittelstadt [16] instead argues that "developers will always be constrained by the institutions that employ them" and that AI ethics must also be the ethics of AI businesses and organizations.

Although these AI ethical codes of conducts are popular, literature suggests that they are insufficient for the development and adoption of responsible AI [2], [9], [11], [13], [43]. The existence of a code of ethics alone does not appear to impact practitioners as McNamara et al. [44] found that simply reviewing the ACM Code of Ethics did not impact the responses of students or professionals to 11 software-related ethical decision scenarios. Translating high-level principles into actionable practices or design fixes remains a major challenge because many principles are ambiguous or challenging to implement [7], [10]. Attempts to codify these standards as technical requirements can result in a "checklist" mentality rather than critical reflective practice [14], [16].

As voluntary and non–legally binding statements, AI ethics codes of conduct have few enforcement mechanisms. There are no legal consequences for deviations from the code of conduct [2]. The most likely consequences for misconduct are reputational loss or restriction on membership in professional societies, but both of these are considered weak enforcement mechanisms [2]. Industry-sponsored initiatives are often marketing tools and in some cases, explicit virtue signaling intended to delay regulation [2], [16]. Hagendorff [2] describes how when institutions adopt their own ethically motivated "self-commitments," it suggests to legislators that self-governance is sufficient and no additional laws are necessary. Jobin [7] suggests that the involvement by the private sector in AI ethics is a "portmanteau to either render a social problem technical or to eschew regulation altogether."

*C. Pre-Deployment Assessments and Audits*

A promising approach for ensuring responsible AI that is gaining traction is pre-deployment assessments and audits. These have been proposed by various stakeholders as key components of the AI development process [8], [18], [45]. Assessments and audits are a means of proactively identifying and mitigating risks to public safety, which can help build public trust in a system and have been widely used in fields like engineering [18], [43], [46]. Algorithmic audits and assessments are already being used in practice [47], [48].

Assessments are non-independent, internal, or second-party evaluations aimed at providing feedback and recommendations to an organization on how an AI tool or algorithm can be improved to align with legal or ethical standards [8]. An AI impact assessment can be an important tool in an AI risk management framework to articulate risks for impact on various stakeholders at the beginning of a development process [40]. Impact assessment frameworks already exist [10], and others are proposed such as a questionnaire that forces developers to consider impacts of a system [11]. An ethical risk assessment explicitly centers the risk that the use of an algorithm negatively impacts the rights and freedoms of stakeholders, rather than legal or compliance risks [8], [45]. UNESCO recommends particular attention of ethical impact assessments on the rights of marginalized and vulnerable people, labor rights, and the environment and ecosystems [45]. A bias assessment can be concurrently conducted with an ethics impact assessment to specifically test an algorithm or AI tool for bias [8], [9]. Assessments can also lead to certifications for particular AI tools, such as the Responsible AI Institute Certification, which is the first accredited certification program of its kind [49].

An audit is an independent assessment or evaluation of an AI tool or algorithm intended to serve society or some other body independent of the evaluated organization, generally providing greater objectivity than an assessment [8]. An audit can take a range of approaches, from checking governance documentation, to reviewing code, to requiring reporting of the accuracy of interpretable modeling methods [47], [50], [51]. Audits have been proposed as a way of creating a certification system for AI technologies [45] or for organizations creating AI tools [12]. Industries like finance have relied for many years on audits as part a "three-lines" defense approach to validate model design, documentation, and deployment plans [40], [52], [53]. Calls for government-mandated AI and algorithm audits highlight the improved safety that can result from independent and transparent auditing procedures [2], [43], [46]. Examples of audits include New York City Council's law requiring bias audits for companies using hiring algorithms [48] and the proposed EU Artificial Intelligence Act requiring audits of algorithms in high-risk contexts [1].

A significant challenge for implementing AI assessments and audits is agreeing and enforcing shared standards [9]. The most progress in this area has been for fully automated AI tools like self-driving cars [54] and for high-risk algorithms [1]. Standards for safety-critical systems like machinery and robotics are already well established [55], and "building codes" in AI development similar to those in architectural design have already been proposed for generic software and medical devices [43]. Efforts like Datasheets for Datasets [26], Good Machine Learning (ML) Practice for medical device software [56], and technical standards for automated driving systems [54] have gained attention. However, despite calls for policymakers to create penalties or incentives for audits [43], all of these are voluntary practices. With only a few exceptions, such as audits of AI hiring tools required in New York [48], the large majority of efforts to codify assessment or audit requirements in U.S. state or federal law have failed because of substantial push back from industry and large projected cost and complexity of implementation [57].

A second major challenge is identifying, educating, and employing the appropriate personnel to complete AI assessments and audits. Financial institutions have a lengthy history of performing internal audits with employed technical staff and third-party audits with appropriate effective oversight [53], [58]. Deloitte, a consulting firm, imagines an algorithmic auditor as one of its "government jobs of the future," anticipating that these individuals will work with regulatory and judicial agencies to review advanced AI algorithms [59]. Individual auditors will likely be insufficient; comprehensive AI assessments or audits will require a team with a variety of backgrounds, including individuals with expertise in philosophy, law, human rights, socio-technical considerations, organizational ethics, statistics, and ML [8], [15], [22]. These teams will require auditing and assessment tools, tests, and methods that have not yet been developed [9], [59]. A concern is that auditors may lack specific technical expertise to complete an audit [9]. Creating a network of AI assessors or auditors and certifying or licensing them may help standardize expertise [45].

The AI assessors and auditors should be a part of a larger governance process to ensure accountability. At organizations, this could include an institutional review board (IRB), a human-subjects protection committee, or a data ethics advisory panel that has the capacity to manage internal AI assessments and systematically identify impact, risks, and approve and reject proposals [9], [11], [18], [60], [61]. Expansion of the purview of IRBs has been considered an especially promising and effective approach [7], [60]. A Chief Model Risk Officer or a Chief AI Ethics Officer can help maintain accountability across an institution and ensure that AI incidents do not occur [8], [51], [52]. National oversight boards, such as a driverless vehicle oversight board, could provide agile, industry-specific standardization, adoption, and accountability for baseline principles, although soft governance approaches like these can be easily overridden or ignored without top-down enforcement [43]. Although all of these governance options could support a more organized approach to ensuring responsible AI, they also come with the monetary and logistical costs of a proliferation of regulatory bodies [43].

*D. Post-Deployment Accountability*

Policies that support post-deployment monitoring and accountability of an AI tool can help build trust and provide redress for harm [9], [18]. A right to explanation, such as the one created by the GDPR, and a formal complaint and appeal process with timely resolution are considered key attributes of organizational commitment to accountability [9], [45]. AI monitoring tools similar to "black box" flight data recorders can produce an audit trail with post-event insight into accidents [43]. Development of insurance schemes, tort law, and other legal frameworks by policymakers can better establish responsibility or culpability [2], [18]. The accountability of users of AI tools, such as medical professionals using AI medical devices or law enforcement officials using facial recognition technology, is also receiving increased attention [43]. End users will require appropriate training on the responsible use of an AI tool [8]. End User License Agreements (EULAs) and Terms of Service (TOS) are legal contracts that specify the rights and restrictions of use. Although these mechanisms have received substantial criticism for violating consumer rights and are sometimes unenforceable [62], licensing frameworks for ensuring responsible AI use have been proposed [63] and a few responsible AI licenses are already available [64].

U.S. agencies and states can play a particularly important role in post-deployment accountability. States are recognized as laboratories of legal innovation, and state-level regulation could prove agile and effective in enforcing responsible AI [65]. In the 2021-2022 legislative session, nine states introduced legislation that would impose some form of oversight on the use of artificial intelligence or algorithms [66]. A single federal agency, or network of agencies, can also facilitate post-release accountability. The Fair Trade Commission (FTC) Act prohibits unfair or deceptive practices, including the sale or use of racially biased algorithms; violation of this can result in FTC enforcement actions [67]. The Equal Opportunity Employment Commission recently released guidance that holds an employer responsible for use of technology (including AI tools) in employment-related decisions that discriminate against individuals with disabilities, even if they do not design or administer the system [68].

Yet, post-accountability policies still have a way to go. Rudin [50] points out that although a "right to explanation" is nice, there are no current requirements in GDPR that an explanation be, "accurate, complete, or faithful to the underlying model." Institutions and processes for filing complaints and appeals should be formalized [69]. Accident investigation for automated AI tools will require both development of black box–like recorders and processes for investigating accidents, neither of which yet exist in a standardized, common format [43]. Decisions related to who holds responsibility for an accident—designer, developer, owner, operator, or overseeing entity—and appropriate compensation are still being determined in the court system [2], [43]. Even when a legal framework emerges, Tutt [65] writes that, "ex post judicial enforcement would likely be too blunt to effectively ensure unsafe algorithms will be kept off the market."

*E. Database of AI Technologies or Incidents*

A database of models is a useful tool for inventorying AI models and supporting transparency. Guidance for banks recommends maintaining a firm-wide inventory of all models and model validations [58]. U.S. Executive Order 13960 [70] requires every agency to annually prepare an inventory of non-classified and non-sensitive use cases of AI, and these can now be found on government agency websites [71]–[73]. The EU is considering required registration of standalone, high-risk AI systems in a public, EU-wide database to increase public transparency and oversight [1].

An incident database can be a helpful tool to collect data on and learn about past failures [52]. When publicly available, incident data can be a valuable resource for researchers, authorities, and developers [43]. Various U.S. agencies have embraced recording of incidents with tools such as the Fatality Analysis Reporting System maintained by the National Highway Transportation Safety Administration [74] and the Adverse Event Reporting System maintained by the Food and Drug Administration (FDA) [75]. Although no government agency currently maintains a public AI incident database, a number of ad hoc initiatives have been started [52] such as AI Incident Database [76], AI Tracker [77], and Awful AI [78]. Although these systems use volunteers to collect publicly available reports, the EU may soon set a new precedent with mandated reporting of high-risk AI incidents [1].

*F. Involvement of Community Stakeholders*

Given the outsized societal impact of AI tools, multiple frameworks call for a focus on community and diverse stakeholder involvement throughout the AI life cycle to support responsible AI development [8], [9], [22], [46]. Participatory stakeholder engagement, including convening of individuals, groups, and community organizations, is recognized as an effective and insightful tool for collecting stakeholder feedback and identifying areas of concern when developing AI tools [22][79]. Organizations are cautioned to not treat participatory engagement as a perfunctory exercise but to instead incorporate engagement throughout the AI development process [14], [22]. Interviewing stakeholders can reveal important concerns and vulnerabilities in an AI tool and can increase the likelihood of identifying problematic assumptions and limitations prior to deployment [8], [9]. Stakeholders may include both subject matter experts and intended users, and the NIST Risk Management Framework suggests that specific attention should be granted to incorporating the views of historically excluded populations, people with disabilities, older people, and those with limited access to the internet [22].

*G. Policies That Support Responsible AI Education*

Implementation of responsible AI practices requires knowledge of responsible AI practices. Integrating responsible AI content into educational offerings is an encouraging approach for growing awareness around the importance of responsible AI [6], [11]. Discussion of data ethics appears to have expanded in university curricula [69], although the number of university AI, ML, or data science courses that mention ethics in the syllabus or course description remains small [80]. Additional evidence suggests that data ethics is more frequently discussed at higher-ranked, elite universities in the United States as compared to lower-ranked U.S. universities or international institutions [17]. Education must also extend beyond university curricula to include partnerships with international organizations and educational institutions to support general AI literacy and reduce digital divides [8], [45], [81]. Continuing education should be available, including ongoing training on how technologies may encode and promulgate bias [46]. A variety of courses and workshops are now available to individuals interested in learning more about responsible AI [82].

*H. Policies That Support Responsible AI Research*

Funding is crucial to research and development, and funding policies can be used to support responsible AI research more directly. As described in Hagendorff [69], "it is no secret that large parts of university AI research are financed by corporate partners." This can result in a conflict of interest because AI research is more closely aligned with corporate goals than public values. UNESCO calls for member states to explicitly support AI ethics research by investing in such research and creating incentives for the public and private sectors to invest [45]. The World Health Organization calls for more research on how ageism, racism, and sexism affect the design and use of AI [46]. Some U.S. government agencies, like FDA, are already investing in research efforts to support regulation and responsible use of AI- and ML-based software [56]. Journals that publish AI and ML research can set standards for responsible AI approaches, such as expanding peer review and requiring consideration of AI risks [83].

*I. Policies That Support Diversity in AI Development*

Diversity among the team members designing, developing, and monitoring AI systems can bring wider perspectives that promote responsible AI development [52]. This includes diversity of experience, expertise, and backgrounds to ensure that AI systems align with a broad group of users [22]. The current lack of diversity within the AI community [69] can result in a small, nonrepresentative group of individuals making significant decisions that may harm a wide community. Policies that promote and increase diversity and inclusiveness can better ensure equal access to AI technologies and their benefits [45]. Organizations and governments can support diverse AI teams through defining policies and hiring practices that facilitate inclusivity, empower contribution of staff feedback without fear of reprisal, and engagement with external expertise where internal expertise is lacking [22].

## III. Discussion

No one policy alone can foster responsible AI, and no one institution can be responsible for implementing responsible AI policies. In this section we consider the use and potential impact of each of the nine policies and suggest how institutions may prioritize investment in each policy.

### A. Use and Potential Impact of Responsible AI Policies

Table I summarizes the relative barriers to widespread use, likelihood of voluntary use, and potential impact of the nine responsible AI policies and practices considered in this analysis. A *Low* barrier to widespread adoption indicates that it would be relatively straightforward for an institution to begin using a policy while a *High* barrier to widespread adoption indicates that substantial development or mechanisms are required for the policy to be widely adopted. A *Low* likelihood of voluntary use suggests that organizations may have a low willingness to adopt a policy if it is not mandated, while a *High* likelihood of voluntary use implies that a substantial number of organizations are likely to implement a policy, even if not mandated. *Low* potential impact denotes policies which, even if implemented, may not substantially contribute to practical application of responsible AI principles, while *High* potential impact signifies policies that could lead to real-world expansion, promotion, and implementation of responsible AI.

TABLE I. USE AND POTENTIAL IMPACT OF RESPONSIBLE AI POLICIES

| Policy | Barriers to Widespread Adoption | Likelihood of Voluntary Use | Potential Impact |
|---|---|---|---|
| Licensure or Certification of AI Developers | High | Low | Medium |
| AI Ethics Statement | Low | High | Low |
| Pre-Deployment Audits or Assessments | High | Medium | High |
| Post-Deployment Accountability | High | Low | High |
| Database of AI Technologies or Incidents | Low | Low | Low |
| Involvement of Community Stakeholders | Medium | Low | Medium |
| Policies That Support Responsible AI Education | Medium | Medium | Medium |
| Policies That Support Responsible AI Research | Medium | High | Medium |
| Policies That Support Diversity in AI Development | Medium | Medium | Medium |

Two policies, *Pre-Deployment Audits or Assessments* and *Post-Deployment Accountability,* are theorized to have *High* potential impact. Both approaches, even when used in few contexts, have shown to have a considerable impact on organizational and AI developer behavior, including modifications to underlying algorithms and alignment with legal standards. Five policies are considered to have *Medium* impact.

*Licensure or Certification for AI Developers* could standardize skillsets and create a system that better holds individuals responsible for development, although as noted previously, individual action alone is not enough to ensure responsible AI development. *Involvement of Community Stakeholders* has been recognized as a key approach to obtaining feedback that can dramatically alter the motivation and development of a project, but similar to licensure or certification, is likely not enough alone to foster responsible AI, especially if it is easy to ignore stakeholder recommendations. *Policies That Support Responsible AI Education*, including for postsecondary students and practitioners, can build awareness of responsible AI best practices. Existing research, including by large AI tech companies, has shown that *Policies That Support Responsible AI Research* can lead to important findings and development of responsible AI tools. Diversity in teams has been shown to be important to inclusive design principles, so *Policies That Support Diversity in AI Development* are also hypothesized to have a medium impact. Two policies, *AI Ethics Statements* and *Database of AI Technologies or Incidents*, are speculated to have *Low* impact; numerous AI ethics statements and databases exist and their impact on responsible AI development is important but small relative to the other policies.

However, these two policies with *Low* potential impact are also the only policies with *Low* barriers to widespread adoption. Many government entities, professional societies, and organizations have already implemented AI ethics statements, and it is relatively straightforward to template new statements from these existing options. Creation of databases for AI technologies or incidents has already garnered interest, and although there are likely arguments to be made for centralizing and standardizing these approaches, some databases already or will likely soon exist. Three policies face *High* barriers to widespread adoption. The first, *Licensure or Certification of AI Developers,* requires not only the creation of responsible AI licensure or certification standards, individual appetite for obtaining a license or certificate, and generating industry demand, but also a clear definition of the AI developer profession and what should be standardized about the role. These challenges are substantial. *Pre-Deployment Audits or Assessments* and *Post-Deployment Accountability* also face *High* barriers to widespread adoption, namely because of the nascency of tools and methodologies to support facilitation, high overhead costs, and the challenges of developing legislation.

The likelihood of voluntary use was estimated with evidence of current or prior use or perceived intent to implement a policy without a mandate. AI ethics statements have already been widely adopted, demonstrating *High* voluntary use. Similarly, there is already demand for funding that supports responsible AI research, and technology companies are already widely using funding to understand and address responsible AI. As previously noted, expanding interest in responsible AI research support to governments and agencies could mitigate current conflict of interest concerns in responsible AI research funding. Organizations have already expressed some interest in pre-deployment audits or assessments, often as part of an internal risk strategy, but the likelihood of voluntary use is rated as *Medium* because these approaches will require considerable investment, which may deter voluntary use. *Policies That*

*Support Responsible AI Education* and *Policies That Support Diversity in AI Development* are both rated as *Medium* likelihood of voluntary use; education institutions and organizations have publicly stated intentions to use these policies, although follow-up is needed to understand if and how they are ultimately applied. *Licensure or Certification of AI Developers* is considered *Low* likelihood of voluntary use given the historic lack of interest from developers and the considerable investment required by many parties to make licensure or certification a success. A narrow licensure or certification mandate, such as for specific development roles, products, or industries, would likely be needed to overcome these barriers. Post-deployment accountability policies will likely see *Low* voluntary use; a legal framework that includes penalties for failure to comply, as utilized by U.S. agencies, will likely be required. A public-facing *Database of AI Technologies or Incidents* would likely see few voluntary contributions by organizations because they generally seek to protect propriety information and do not want negative press for AI incidents. Mandates, such as the executive order requiring inventorying of AI tools, appear to be more effective than voluntary use. Finally, involvement of community stakeholders is currently low in AI development and future voluntary inclusion of stakeholders is likely to continue to be *Low* without mandated changes.

The ideal scenario would likely be a policy with low barriers to adoption, high likelihood of voluntary use, and high potential impact. No policy matches this scenario, so tradeoffs must be made. The highest impact will likely be from *Pre-Deployment Audits or Assessments* and *Post-Deployment Accountability*, but these policies also have the largest barriers to adoption. *AI Ethics Statements* and *Database of AI Technologies or Incidents* are much easier to implement but are unlikely to have substantial impact and in the case of *Database of AI Technologies or Incidents,* are unlikely to see considerable growth with voluntary use alone.

### B. Recommendations for Prioritization by Institution Type

Multiple policy approaches will be necessary to ensure development and implementation of responsible AI, and different institutional stakeholders can focus their limited resources for targeted policy efforts most closely associated with their spheres of influence. Table II suggests a prioritization of responsible AI policies for the U.S. national legislature, U.S. government agencies, U.S. states, professional societies, and organizations, including both public and private sector companies. Prioritization is based on the findings in Table I and the applicability of a given policy to the particular institution type.

The U.S. Congress should highly prioritize investment in post-deployment accountability. This could include development and implementation of a legal framework, perhaps with similarities to the EU Artificial Intelligence Act, that sets standards for monitoring and auditing AI technologies and providing recourse and appeals for AI harms. The government-mandated creation of inventories of non-confidential government AI technologies by agencies was effective; this should continue as a priority. The national legislature should increase investment in policies that support responsible AI education and AI research, such as allocating additional funding for these efforts.

TABLE II. PRIORITIZATION RECOMMENDATIONS

| Policy | U.S. Legislature | U.S. Government Agencies | U.S. States | Professional Societies | Organization |
|---|---|---|---|---|---|
| Licensure or Certification of AI Developers | ⊘ | ✓ | ⊘ | ✓+ | ✓- |
| AI Ethics Statement | ✓- | ✓- | ⊘ | ✓ | ✓- |
| Pre-Deployment Audits or Assessments | ⊘ | ✓+ | ⊘ | ✓ | ✓+ |
| Post-Deployment Accountability | ✓+ | ✓ | ✓ | ⊘ | ✓ |
| Database of AI Technologies or Incidents | ✓ | ✓ | ⊘ | ⊘ | ✓- |
| Involvement of Community Stakeholders | ⊘ | ✓ | ⊘ | ⊘ | ✓+ |
| Policies That Support Responsible AI Education | ✓ | ⊘ | ✓+ | ✓ | ✓ |
| Policies That Support Responsible AI Research | ✓ | ✓+ | ⊘ | ✓+ | ✓- |
| Policies That Support Diversity in AI Development | ⊘ | ✓ | ⊘ | ✓ | ✓+ |

⊘ = No investment   ✓- = Low Priority
✓ = Priority   ✓+ = High Priority

Nearly all of the proposed policies could apply to U.S. government agencies. They can play a particularly important role in leading policies related to *Pre-Deployment Audits or Assessments.* Although it seems like individual agencies are exhibiting varying approaches to this, such as NHSTA considering regulations of driverless vehicles while FDA regulates AI in medical devices, U.S. government agencies should heed calls for a new singular governing committee that can holistically review and regulate AI technologies [65]. U.S. government agencies should also prioritize funding for responsible AI tools and methods and require commitment to responsible AI tenets in AI contracts. These funding and contract stipulations can include requirements, such as considering diversity in AI development and involvement of stakeholders. Government agencies themselves can also lead focus groups and listening sessions with community stakeholders for citizen input. U.S. government agencies are already required to maintain inventories of their non-confidential AI technologies; a centralized database of these tools could be an even more accessible resource. Agencies can be actively involved in facilitating standardization and requirements for licensure or certification of AI developers, such as a certificate required to deploy a specific AI technology. Agencies including the FTC are already involved in post-deployment accountability; this work should be expanded across sectors. The lowest priority for agencies is AI ethics statements. Some agencies have already made published commitments regarding AI ethics [56], [84], and although these are useful for setting shared values, agencies can be more effective with other policies.

U.S. states have a more limited set of suggested responsible AI policy priorities. A major reason for this is that it would be frustrating to have a patchwork of certifications, pre-release audits, or databases in a world where AI technologies readily transcend borders. A high priority for states is policies that support responsible AI education. K-12 standardized curricula and associated tools are almost always developed at the state level [85], and given the ubiquity in use of AI tools, there is a strong case to be made for improving education related to the use of AI in youth education. States can also encourage AI curricula development in public postsecondary institutions, including creation of responsible AI research centers. Another priority for states is post-deployment accountability. States have historically been strong venues for prototyping laws later implemented nationally. Although a patchwork of policies is not ideal, a state-based legal framework could provide incentive for organizations to adopt responsible AI approaches without waiting for national regulations.

Professional societies, including but not limited to ACM and IEEE, can take the lead in licensure or certification of AI developers. These efforts should carefully note the failures of past similar efforts and develop a strategy for a clear and narrowly focused certification that is developed in collaboration with U.S. government agencies and includes input from public and private sector organizations. Professional societies often support many research activities, so research policies are a high priority. At conferences and in publications, professional societies can arrange presentation tracks, workshops, special sessions, and roundtables with a specific focus on responsible AI. This type of collaboration facilitates ideating and networking that can support larger policy efforts. Professional society journals should also require AI research contributions to meet responsible AI standards, such as diverse data and stakeholder input. Other priorities including releasing an AI ethics statement, which many professional societies have already done, and including member organizations to commit to AI ethics principles. Professional societies should be directly involved in collaborating with U.S. agencies on development of pre-deployment audit and assessment standards. Professional societies can support continued education of members through webinars and provide development pipelines for students with a focus on responsible AI. To support increased diversity in AI development, professional societies can amplify the work of diverse AI researchers through internal and external communication channels, providing networking and mentoring spaces for underrepresented members, and generally promote the need for increased diversity in AI development.

Finally, organizations, including public and private sector companies, will need to consider facilitating nearly all of the policy options here. This will be a heavy lift, so organizations should first prioritize pre-deployment audits and assessments because this is one of the policy approaches with *High* likely impact. Two crucial components of this approach are employing staff with appropriate audit or assessment expertise, including educating existing staff, hiring new staff, or partnering with external organizations, and development of internal governance mechanisms like an Algorithm Review Board or Chief Ethics Officer. A second high priority of organizations is facilitating community involvement in the AI development process. This will help ensure that AI tools are aligned with community needs and anticipate potential harms long before they occur. A third high priority is implementing policies that support diversity in AI development. This focus aligns well with recent corporate initiatives to increase diversity and inclusion in hiring. A general priority for organizations is to contribute to post-deployment accountability initiatives such as using EULAs or TOS to ensure user accountability for responsible use of an AI technology. Organizations can support internal responsible AI education for their own employees. Licensure or certification development is a lower priority for organizations, noting that should licensure or certification be successful, organizations will likely be interested in including these in job postings or upskilling employees. AI ethics statements are useful for organizations to publicize their commitment to AI ethics and agree on a shared AI ethics framework but are unlikely to lead to direct impact alone. Organizations may someday be mandated to report AI technologies or incidents to a legislative body, but in the short term, the only database prioritization is likely for an internal AI inventory to support AI risk management best practices. Finally, organizations including private companies can continue to financially support responsible AI research but given the well-publicized conflicts of interest in this space, limited resources may be better prioritized elsewhere.

## IV. Conclusion

We have reviewed nine policy options proposed to further responsible AI, considered the likely use and impact of these approaches, and made recommendations for prioritizing implementation across five types of institutions. No one approach alone will ensure responsible AI, and greater alignment can facilitate a more cohesive effort. We encourage stakeholders to use this as a resource to inform policy strategy and development.